\documentclass[twocolumn,prl,aps,showpacs]{revtex4}
\usepackage{xspace,amsmath,amsfonts,amsthm,amssymb,amsbsy,graphicx,color}

\begin{document}

\newtheorem{theo}{Theorem}
\newtheorem{lemma}{Lemma} 

\title{Engineering Quantum States of a Nano-Resonator via a Simple Auxiliary System}

\author{Kurt Jacobs}

\affiliation{Department of Physics, University of Massachusetts at Boston, 100 Morrissey Blvd, Boston, MA 02125, USA}

\begin{abstract}
We show how to engineer an extensive range of non-linear Hamiltonians for a nano-mechanical resonator. The technique requires only a time dependent drive applied to a Cooper-pair box or second oscillator to which the nano-resonator is coupled. This method allows one to generate a large number of non-classical states, as well as Hamiltonians whose classical counterparts are chaotic. 
\end{abstract} 

\maketitle

Nano-mechanical resonators can now be built with frequencies of $100 \mbox{MHz}$ and quality factors of $10^5$~\cite{Blencowe04}. Coupled with recent advances in the measurement~\cite{LaHaye04} and cooling~\cite{Hopkins03,Wilson04,Zhang05,Kleckner06,Naik06} of these systems, this has opened up the exciting prospect of observing quantum behavior in mesoscopic mechanical systems for the first time. To fully explore the quantum behavior of nano-resonators, and consider exploiting this behavior in future applications, one must prepare these systems in highly non-classical states, which in turn requires the action of a non-linear Hamiltonian. A number of approaches have been taken to this problem. One is to engineer non-linear couplings to~\cite{Xue07}, or engineer a reservoir via~\cite{Rabl04}, a Cooper-pair Box (CPB), but so far only a very limited set of Hamiltonians have been obtained in this way. A second approach is to use measurement and/or feedback~\cite{Jacobs07b,Martin07}. While this has potential, low-noise measurements are, at least presently, much more difficult to perform than Hamiltonian control. A third more exotic approach borrows from quantum optics, and involves embedding a quantum dot on the resonator, which in turn interacts with laser fields~\cite{Wilson04}.  Here we show that a large range of non-linear Hamiltonians, and a correspondingly large range of non-classical states, can be engineered by coupling the resonator to a single CPB~\cite{Makhlin01, Irish05,Wei06} or second nano-resonator. This  does not require dissipative, measurement or feedback processes, but simply the application of time-dependent control fields to the auxiliary system, which is a relatively straightforward process. Note that our purpose here is to engineer nonclassical states of the resonator alone, and not an entangled state of the coupled systems. The latter can be achieved with a simple linear Hamiltonian, as shown in~\cite{Armour02,Tian05}.  

This work was inspired by a 2004 article of Lloyd, Landahl and Slotine~\cite{Lloyd04}, in which they point out that, at least in principle, a single interaction of any system with a second simple system should suffice to create any dynamics between the two systems, so long as the interaction operator, $G$, does not commute with the system's Hamiltonian, $H$. This is because commutators of $G$ and $H$ will in general generate a complex algebra. These authors did not give a method for achieving any specific  evolution, however, and there is probably no general answer. For the problem of state-preparation, the task is complicated by the fact that the interaction will in general entangle the two systems. To prepare a state of the primary system, we must entangle and then disentangle them in such a way as to leave a non-trivial operation on the primary system. One is reminded of attempting to flip an edge piece in Rubics cube while leaving the other facets intact. 

Here we show how to generate a significant range of non-linear Hamiltonians for a nano resonator by performing time-dependent rotations on a single qubit to which it is coupled. 
We note that we are not the first to show that general quantum states of an oscillator can be engineered by time dependent driving of a low-dimensional auxiliary system. A scheme to do this was first proposed by Law and Eberly~\cite{Law96}, and another more recently by Santos~\cite{Santos05}. These methods should also be realizable in nano-resonators, although they are very different from the technique presented here; these schemes are most appropriate for creating arbitrary states with small numbers of phonons, since they become increasingly complex as the number of phonons increases. Conversely, the method described here is most appropriate for creating states of arbitrarily large phonon number, such as mesoscopic Scr\"{o}dinger-cat states, and states generated by chaotic systems such as the Duffing oscillator~\cite{Bhattacharya00}. Further, our approach is not limited to two-level auxiliary systems. We show how to generate many non-linear Hamiltonians with an auxiliary oscillator, which may well be important in nano-mechanical applications, and the method could certainly be extended to other systems. 

We consider first a Cooper-pair Box as the auxiliary system. If we electrically bias the nano-resonator and place it adjacent to a CPB, then the two systems interact via the Hamiltonian $H_1 = \alpha x\sigma_z$, where $\alpha$ is the coupling constant determined by the geometry and bias voltage, $x = a + a^\dagger$ is the position of the resonator and $\sigma_z$ is the Pauli operator for the CPB qubit in the charge basis. If we wish to move into the interaction picture to eliminate the free rotation of the resonator, we can preserve this time independent interaction by modulating the coupling strength at the resonator frequency and making the rotating wave approximation. If instead we far-detune the CPB from the resonator, and operate the CPB at the charge degeneracy point, then the interaction becomes $H_2 = \alpha' N \sigma_x$ where $N\equiv a^\dagger a$~\cite{Sarovar05}. We can obtain an effective oscillator Hamiltonian proportional to $N$ either by changing the modulation frequency of the coupling strength, or by employing $H_2$ with the qubit in a $\sigma_x$ eigenstate. Finally, we can perform arbitrary rotations of the CPB qubit by changing an adjacent gate voltage~\cite{Makhlin01}. These are the basic interactions and operations that we wish to use to create non-linear Hamiltonians for the resonator. We will do this by using the interactions, along with sequences of rapid rotations applied to the qubit. In what follows we set $\hbar =1$ to simplify the notation.

The first tool we employ is the following: given the interaction Hamiltonian $A\sigma_j$, where $A$ is an operator of the resonator and $\sigma_j$ is a Pauli operator for the qubit, one can generate the evolution $e^{-i A \mathbf{n}\cdot\boldsymbol{\sigma}}$. To do so one first applies a rapid rotation to the qubit given by $e^{i\theta \mathbf{m}\cdot\boldsymbol{\sigma}}$ (where $\mathbf{m}$ is defined by $[\mathbf{m}\times\mathbf{n}]\cdot\boldsymbol{\sigma} = \sigma_j$), allows the systems to interact, and then applies the inverse rotation. The result is the desired unitary, since $e^{-i\theta \mathbf{m}\cdot\boldsymbol{\sigma}} e^{ A \sigma_j} e^{i\theta \mathbf{m}\cdot\boldsymbol{\sigma}} = e^{-i A \mathbf{n}\cdot\boldsymbol{\sigma}}$. 

Our second tool is the Zassenhaus formula, which may be regarded as complementary to the more well-known BCH-formula~\cite{Magnus54,Richtmyer65}, and allows one to write an exponential of the sum of two operators as an infinite product of exponentials of the operators and their commutators: $e^{ (A+B)} = e^{ A}e^{ B}e^{C_2}e^{C_3}\cdots$~\cite{Witschel75}. The second and third order terms in the product expansion are 
\begin{eqnarray}
   C_2 & = & - (1/2) [A,B] , \\ 
   C_3 & = & (1/3) [B,[A,B]] + (1/6) [A,[A,B]] ,
\end{eqnarray}
and further terms are given by a recursion relation~\cite{Wilcox67}. 

We now assume (without loss of generality) that the interaction between the resonator and auxiliary qubit is $\alpha B\sigma_x$, and use the above procedure, in tandem with the Zassenhaus formula, to generate an evolution proportional to the square of $B$. To do so we first change the qubit representation to $\tilde{\boldsymbol{\sigma}} = (\tilde{\sigma}_x,\tilde{\sigma}_y,\tilde{\sigma}_z) = ([\sigma_x + \sigma_z]/\sqrt{2},\sigma_y,[\sigma_x - \sigma_z]/\sqrt{2})$. The evolution due to the interaction during a time interval $\Delta t$ becomes $\exp(i A(\tilde{\sigma}_x + \tilde{\sigma}_z))$, were we have defined $A = \alpha \Delta t B$ to simplify the notation. We then expand this expression to second order using the Zassenhaus formula, and multiply on the left by $e^{-i A \tilde{\sigma}_z} e^{-i A \tilde{\sigma_x} }$ to obtain  
\begin{equation}
   e^{-i A \tilde{\sigma}_z} e^{-i A \tilde{\sigma_x} } e^{i  A\sigma_x} =   e^{-i A^2 \sigma_y/2} , 
   \label{eq::seq1} 
\end{equation} 
valid to second order in $A$. This gives us the following procedure for generating the Hamiltonian $A^2$. We first apply control rotations to the qubit to prepare it in an eigenstate of $\sigma_y$. We then wait a sufficiently short time interval during which the systems interact via $\alpha A\sigma_x$. To complete the operation we apply the unitaries $e^{-i A \tilde{\sigma}_x}$ and $e^{-i A \tilde{\sigma}_z}$, by using the procedure described above. During this sequence the two systems are entangled and then disentangled, and the resonator undergoes the non-trivial evolution $e^{\pm i (\alpha \Delta t)^2 B^2}$. Here the sign is given by the choice of the initial qubit eigenstate, and the evolution time must be small compared to the eigenvalues of $\alpha B$. Applying this full sequence repeatedly then generates this evolution for an arbitrary time $t$. 

Note that using the above procedure to generate the evolution $A^2$ requires that the time interval in each repetition be sufficiently small that the third and higher order terms in the Zassenhaus expansion can be ignored. We can therefore increase dramatically the procedure's efficiency by obtaining an expression similar to Eq.(\ref{eq::seq1}) that is valid to third order instead of second. We can do this by noting that if we flip the signs of both $A$ and $B$ in the Zassenhaus formula, the sign of the second order term does not change, but that of the third order term does. We can therefore cancel the third order term by applying the sequence again with the signs flipped. This gives 
\begin{equation}
   e^{i A \tilde{\sigma}_z} e^{i A \tilde{\sigma_x} } e^{-i  A\sigma_z}e^{-i A \tilde{\sigma}_x} e^{-i A \tilde{\sigma_x} } e^{i  A\sigma_x} =   e^{-i A^2 \sigma_y/2} , 
\end{equation}
which is now valid to third order in $A$. 

In creating the effective Hamiltonian $A^2$, we have also obtained the effective interaction $A^2 \sigma_y$. We can therefore apply the above procedure to this new interaction to obtain the effective Hamiltonian $A^4$. In theory this procedure allows us to generate all even powers of $A$.  

\begin{figure*}[t] 
\leavevmode\includegraphics[width=0.9\hsize]{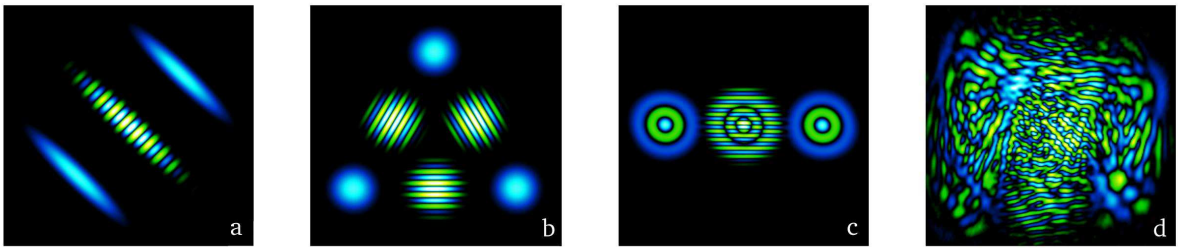}
\caption{Wigner functions for various states engineered with the techniques discussed. Plot key: luminosity is proportional to absolute value, with blue positive and green negative. (a) A squeezed Schr\"{o}dinger-cat state; (b) three superposed coherent states; (c) two superposed displaced number states; (d) a state generated by the Duffing oscillator after ten drive periods.} 
\label{fig1}
\end{figure*}

To create odd powers of an operator $A$, we can use the same method, but this time expand the Zassenhaus formula to third order.  This gives 
\begin{equation}
   e^{i A^2 \tilde{\sigma}_y /2}  e^{i A \tilde{\sigma}_z} e^{i A \tilde{\sigma_x} } e^{ - i  A\sigma_x} =   e^{- i  A^3 [2\tilde{\sigma}_x - \tilde{\sigma_z}]/12} . 
   \label{eq::seq2}
\end{equation} 
To apply the first unitary on the left hand side, we can use essentially Eq.(\ref{eq::seq1}), but we must change the sequence on the left hand side a little so as to flip the sign on the right hand side. The operation we require is 
\begin{equation}
   e^{i A \tilde{\sigma}_z} e^{-i A \tilde{\sigma_x} } e^{i  A\sigma_z} =   e^{i A^2 \tilde{\sigma_y}/2} e^{- i  A^3 [2\tilde{\sigma}_x + \tilde{\sigma_z}]/12} . 
   \label{eq::seq3}
\end{equation} 
Combining Eq.(\ref{eq::seq2}) and Eq.(\ref{eq::seq3}) we obtain finally
 \begin{equation}
   e^{i A \tilde{\sigma}_z} e^{-i A \tilde{\sigma_x} } e^{i  A\sigma_z} e^{i A \tilde{\sigma}_z} e^{i A \tilde{\sigma_x} } e^{ - i  A\sigma_x}=  e^{- i A^3 \tilde{\sigma_z}/6} , 
   \label{eq::seq4} 
\end{equation} 
which is an explicit pulse-sequnce for generating $A^3$. 
We can also increase the efficiency of this procedure, as we did for the $A^2$ sequence above, by expanding the Zassenhaus formula to forth order, and choosing a sequence that eliminates the forth order terms. We now have explicit methods to generate second, third and forth powers of any interaction operator,  and it is clear that by expanding the Zassenhaus formula to higher orders, we can generate higher powers if desired. 

As discussed above, we have at our disposal interaction Hamiltonians with $A \propto x$ and $A \propto N$. By using both together we can create all real linear combinations of powers of $x$ and $N$. Note that the squeezing Hamiltonian is in this set, being $a^2 + a^{\dagger 2} \equiv x^2 - 2 N$. To apply both interactions together (or sequentially, as these are equivalent to first order) to obtain the desired linear combinations, the simplest method would probably be to have the resonator interact with two CPB's, one on-resonance and the other detuned. 

So far we have only used time-dependent control performed on the qubit. It is worth noting that if we also apply rapid ``rotations'' to the resonator, given by $e^{i \theta N}$, then we can generate an effective interaction proportional to a linear combination of $x$ and $p$ by using the relation $e^{-i \theta N} e^{i\lambda x \sigma_z} e^{i \theta N} = e^{i [\cos(\theta)x + \sin(\theta) p] \sigma_z}$. This allows us to further extend the accessible set of non-linear Hamiltonians.  Rapid rotations of the resonator are equivalent to shifting the resonator frequency, which could be achieved by using the interaction $\lambda N \sigma_z$, and modulating $\lambda$. 

We now consider using an oscillator as the auxiliary system. Coupling to an auxiliary nano-resonator via $\lambda x X$ is straightforward~\cite{Cleland04}, where $X\equiv (b + b^\dagger)$ is the position of this resonator. Using a superconducting L-C oscillator as the auxiliary system one can obtain the interaction $\mu N b^\dagger b$~\cite{Buks06}. We allow ourselves two simple control operations on the auxiliary oscillator. The first is a linear drive, and the second is a frequency shift (phase-space rotation).  Given an interaction $\alpha A(b + b^\dagger)$, where as before $A$ is a resonator operator, the rotations allow us to obtain all effective interactions of the form $\alpha A(be^{-i\theta} + b^\dagger e^{-i\theta})$. The Zassenhaus formula provides the exact relation 
\begin{equation}
   e^{i A X} e^{i A P} e^{-i A (X+P)} =  e^{- i A^2 } , 
\end{equation} 
giving a pulse sequence to generate $A^2$. Because this relation is {\em exact}, the pulse sequence need only be applied {\em once} to obtain $A^2$, in contrast to the sequences above. This advantage follows because the Zassenhaus sequence $C_n$  terminates for $X$ and $P$. As a result, we cannot use $X$ and $P$ to obtain higher powers of $A$. 

We can obtain all even powers of $N$ by using $e^{-i \lambda N b^\dagger b + i\alpha b + i\alpha^* b^\dagger}$ and expanding as before with the Zassenhaus formula. Odd powers can be obtained by sandwiching $e^{-i N b^\dagger b}$ between linear driving pulses $e^{\pm i\alpha X}$ to give $e^{-i N b^\dagger b + i \alpha N P}$, and once again applying Zassenhaus. As an example, the sequence we obtain to generate $N^3$ (valid to third order) is $ U(-\lambda) U(\lambda) = e^{i (2/3) \lambda^2 N^3 } $ where 
\begin{equation*} 
   U(\lambda) = e^{i \lambda N P} e^{i N b^\dagger b} e^{-i N (b^\dagger b + \lambda P)} 
\end{equation*}

We can use the above methods to engineer a range of non-classical states, including: 

{\em Squeezed Schr\"{o}dinger-cat states:} A ``Schr\"{o}dinger Cat'' is a superposition of two different coherent states. To generate a superposition of two squeezed states, we first cool the resonator to the ground state, and then apply the squeezing Hamiltonian $a^2 + a^{\dagger 2}$ to produce a squeezed vacuum state. Next we apply a classical linear drive, given by the Hamiltonian $\alpha a + \alpha^* a^\dagger$ to shift the state  in phase space. Finally we apply the Hamiltonian $\lambda (a^\dagger a)^2$ for a time $t = \pi/(2\lambda)$, and this generates a superposition of two squeezed states symmetrically placed about the origin~\cite{Mancini97,Bose97}. The Wigner function for such a state is displayed in Fig. 1 (a).   

{\em Multi-cat states:} To generate a superposition of more than two coherent states we can again use the Hamiltonian $\lambda (a^\dagger a)^2$. Applying this Hamiltonian to an initial coherent state for times $t=\pi/(3\lambda)$ and $t=\pi/(4\lambda)$ creates superpositions of three and four coherent states respectively. The Wigner function for the former is shown in Fig. 1 (b). 

{\em Displaced-number cat states:} A displaced number state is a number state that has been shifted in phase space~\cite{Lvovsky02}. We can create number states in a straightforward way by starting with the ground state, and using the qubit to successively add single phonons to the resonator. To do this we employ the interaction $\lambda x \sigma_x$, which under the rotating wave approximation simply generates an exchange of excitation between the qubit and the resonator. If we start the qubit in its ``upper state'' $|1\rangle$, and the resonator in the ground state, $|0\rangle$, then after a time $\pi/\lambda$ the resonator is in state $|1\rangle$. Repeating this one can obtain any $|n\rangle$. We can then displace these number states by applying a coherent drive, and subsequently form a superposition of two or more by applying $\lambda (a^\dagger a)^2$. A superposition of two displaced number states with $n=2$ is  shown in Fig. 1 (c).  

Preparing non-classical states is not the only motivation for generating non-linear Hamiltonians in nano-electro-mechanical systems. Such Hamiltonians also allow us to engineer specific kinds of {\em dynamics} for these systems, including those whose classical counterparts are chaotic. These dynamical systems are of special interest from the point of view of the quantum-to-classical transition, and especially the emergence of classical chaos. This is of particular interest for nano-resonators, because 1) it has been shown that the continuous measurement of the position of a non-linear resonator is sufficient to induce the transition from quantum dynamics to classical chaos~\cite{Bhattacharya00,Habib06}, and 2) a continuous position measurement of a nano-resonator has already been realized experimentally~\cite{LaHaye04}. An example of a chaotic system with a single degree of freedom is provided by the Duffing oscillator, whose Hamiltonian is $\alpha a^\dagger a - \lambda x^2 + \mu x^4 + \Lambda \cos(\omega t) x$~\cite{Bhattacharya00}. This Hamiltonian can be engineered using the above techniques, since we can generate $a^\dagger a$ by detuning, $x^2$ and $x^4$ using the methods above, and the final term is merely a time dependent linear drive.  As an example, we illustrate a typical state generated by the Duffing oscillator after $10$ periods of the drive, from an initial coherent state, and using the parameters in~\cite{Bhattacharya00} with $\hbar=1$. 

We now discuss realizing the above techniques with current experimental parameters. Our goal is to generate dynamics that is significantly faster than the damping rate of a nano-mechanical resonator, which is currently about $\Gamma = 10^{4} \mbox{s}^{-1}$. The control operations are applied to the auxiliary system (CPB or auxiliary oscillator) by switching the voltage on an adjacent gate. It is not difficult to achieve switching times on the order of a nano-second, and significantly faster switching can be achieved with advanced electronics.  Since the dynamics engineered in the resonator must be smaller than the switching rate, this is a limiting timescale. 

Consider as a concrete example preparing the cat state $(|\alpha\rangle + |\!\!-\!\!\alpha\rangle)/\sqrt{2}$, with $\langle \alpha | N | \alpha\rangle = 10$. To do this using an L-C oscillator we require the interaction $\mu N b^\dagger b$, and in this case $\mu$ can realistically be as high as $10^8 \mbox{s}^{-1}$~\cite{Buks06}. (For a CPB, the equivalent interaction strength can be at least $10^6 \mbox{s}^{-1}$~\cite{Jacobs07b}.) If we set $\mu = 2.5\times 10^{6} \mbox{s}^{-1}$, and generate the Hamiltonian using 50 third-order pulse sequences, so as to limit unwanted higher-order terms to $2\%$, we can prepare the state in a time of $\tau = \sqrt{50} \times 2\pi/(4\mu) = 5 \mbox{$\mu$s}$. A switching time of $1 \mbox{ns}$ should be adequate for this purpose, as the whole procedure will require a few hundred pulses. (With faster electronics we could use more pulses and achieve higher accuracy.) To gauge the effect of decoherence, we now simulate numerically the action of the $N^2$ Hamiltonian with the addition of thermal damping at $\Gamma = 10^{4} \mbox{s}^{-1}$ and a temperature of $10 \mbox{mK}$. For a $100 \mbox{MHz}$ resonator we find that the damping reduces the height of the interference fringes from that of a perfect cat by about $50\%$. This would therefore be an appropriate scenario for probing environmental decoherence. The L-C oscillator is also subject to decoherence at a similar rate, but we expect this to have less effect on the final state, since it is repeatedly disentangled from the resonator during the pulse sequence. On the other hand, decoherence times for CPB's are considerably shorter. While these have been steadily improving, they currently stand at about $1~\mu\mbox{s}$~\cite{Gambetta06}, and this might well  interfere with the preparation process.  A detailed investigation of the effects of decoherence in the auxiliary system will be the subject of future work. 


  \end{document}